\documentclass[pdftex,superscriptaddress,aps,onecolumn]{revtex4}
\usepackage{multirow,enumerate,epsfig,graphics,amssymb,amsmath,subeqnarray,mathrsfs,color,xcolor}
\usepackage{color,epsfig,graphics,amssymb,amsmath,subeqnarray,graphicx,amsthm,subfigure,mathrsfs}
\usepackage[colorlinks=true, citecolor=red, linkcolor=blue ]{hyperref}

\newcommand{\change}[1]{{\color{black} #1}}

\newcommand{\dd}{\mathrm{d}}
\newcommand{\ee}{\mathrm{e}}\newcommand{\ci}{\mathrm{i}}

\newcommand{\ub}{\mathbf{u}}

\newcommand{\xb}{\mathbf{x}}

\newcommand{\pard}[2]{\frac{\partial #1}{\partial #2}}

\makeatletter
\def\sgn{\mathop{\operator@font sgn}}
\makeatother
\allowdisplaybreaks[1]

\begin{document}
\title{Universal optimal geometry of minimal phoretic pumps}
\author{S\'ebastien Michelin}
\email{sebastien.michelin@ladhyx.polytechnique.fr}
\affiliation{LadHyX -- D\'epartement de M\'ecanique, Ecole Polytechnique -- CNRS, Institut Polytechnique de Paris, 91128 Palaiseau, France.}
\author{Eric Lauga} 
\email{e.lauga@damtp.cam.ac.uk}
\affiliation{Department of Applied Mathematics and Theoretical Physics, University of Cambridge, Cambridge CB3 0WA, United Kingdom.}
\date{\today}

\begin{abstract}
Unlike pressure-driven flows, surface-mediated phoretic flows provide efficient means to drive fluid motion on very small scales. Colloidal particles covered with chemically-active patches with nonzero phoretic mobility (e.g. Janus particles) swim using self-generated gradients, and similar physics can be exploited to create phoretic pumps. Here we analyse in detail the design principles of phoretic pumps and show that for a minimal phoretic pump, consisting of 3 distinct chemical patches, the optimal arrangement of the patches maximizing the flow rate is universal and independent of chemistry.
\end{abstract}

\maketitle

\section{Introduction}

The  rapid development of  microfluidics, which   already has had a deep impact on both biology and chemistry~\cite{whitesides01,beebe02,hansen03,whitesides2006},  was enabled by key advances in continuum physics. Indeed, it is our understanding of the surface-dominated physics at the micron scale which has allowed the invention of   a whole array of small-scale devices  to precisely control flow and transport processes in microfluidic devices~\cite{squires2005}.  

One of the standard issues in small devices is the difficulty of driving    flows.  In a straight channel, or pipe,   the   rate at which a Newtonian liquid flows  from one side of the channel to the other scales  as the fourth power of a relevant cross-sectional channel length scale times the applied external pressure gradient~\cite{stonereview}. When length scales become tens of microns or less, the   external pressures required to drive flows become prohibitively large and as a result the community has turned to  surface-driven methods where a flow is induced locally~\cite{squires2005,stonereview}. 

In the biological world, surface flows are often created along tissues, or groups of cells, by the  time-varying beating of short cilia~\cite{sleigh1988} resulting in   effective slip boundary conditions for the neighbouring  flow~\cite{Blake1971a,brennen1977}. Although artificial cilia  have been realised in the lab, the dynamics and performance of biological ciliary arrays  has proven difficult to reproduce experimentally~\cite{fahrni2009,babataheri2011,coq2011}, 
 
Instead, a popular method to generate flows near surfaces in the lab   consists in taking advantage of  phoretic mechanisms where externally-applied physico-chemical gradients (such as charge, temperature, composition...) create local body forces on the fluid \change{in thin layers} near surfaces which, through the action of viscous stresses, entrain a bulk flow~\cite{anderson1989}. 
 A famous  example  of such methods is electrophoresis wherein an electric field applied along a channel  filled with an electrolyte drives a flow due to charge imbalance near the electrical double layer at the junction between the fluid and  surfaces~\cite{squires2005}.

While externally-applied gradients are able to drive flows, gradients  which are instead generated locally on the surface of   colloidal particles  can be used to generate locomotion~\cite{Howse2007,Ebbens2011,julicher2009,moran2017}.  
Self-propulsion of such  phoretic   swimmers   can result  either from chemical gradients  directly patterned on the particles themselves via coated catalysts~\cite{golestanian2007}
or   from transport  instabilities for chemically-homogenous  particles \cite{thutupalli2011,michelin2013c,izri2014}, and have proven popular model systems in the field of 
active matter~\cite{marchetti_review}. \change{A canonical example of such catalytic reactions is the decomposition of hydrogene peroxide on platinuum-coated surfaces~\cite{Howse2007} or iron oxide catalysts~\cite{palacci2013}, but many other chemical reactions have also been considered~\cite{duan2015,yadav2015}}.

 The   physico-chemical principles used for phoretic swimmers can in principle also be exploited to induce flow transport in confined devices such as microchannels, and therefore to create pumps~\cite{michelin2015b,shen2016,yang2016,tan2017}. {Yet, the existing literature has only so far provided limited insight on the fundamental design principles of such pumps, and we propose \change{here} a detailed analysis of the link between pump design and performance. In particular, }with a view toward experimental realisation, an important practical question is that of  minimal geometrical design. What type of surface chemistry would be  simple to fabricate yet effective at creating transport? 

In the case of swimmers, the minimal design is that of a Janus particle  whose   surface  is covered by  two distinct, homogeneous  patches of which at least one is phoretically active. By symmetry, a Janus channel cannot be used to pump flows, and the simplest design has three patches. In this \change{paper}, we solve theoretically the $P$-patch problem. We demonstrate that in the minimal case of $P=3$ patches, the optimal pump design, i.e.~the geometrical arrangement of chemical patches leading to maximum phoretic  flow rate, is universal and independent of chemistry, in stark contrast with   Janus particles.

\begin{figure}[t]
\begin{center}
\includegraphics[width=0.55\textwidth]{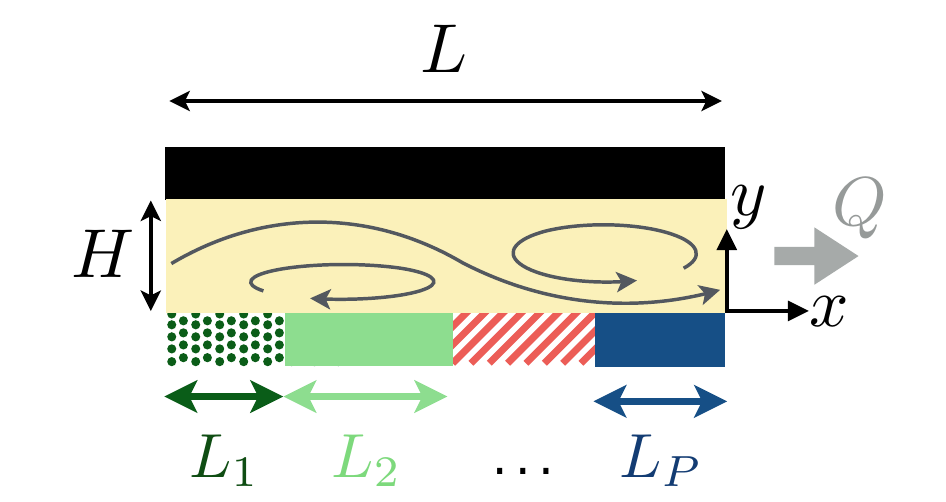}
\caption{\textbf{Periodic phoretic pump design:}
A  straight, two-dimensional channel of width $H$ is periodically coated    with $P$  chemically-active patches (per period $L$) of lengths $L_1,L_2,...,L_P$ (transverse stripes in three dimensions).  
Diffusiophoresis leads to   pumping with flow rate $Q$ (\change{schematic flow illustration}). 
}
\label{fig1}
\end{center}
\end{figure}

\section{Results}
\subsection{Model and performance of a generic phoretic pump}
 
 We consider an infinite,  straight two-dimensional  channel of width $H$ (Fig.~\ref{fig1}). One of the channel walls, located at $y=0$,   is chemically-coated with a catalyst     along a repeated pattern of  period $L$.   The catalyst allows a
 chemical reactant in the liquid  to   produce a new solute species of concentration $C(\xb)$. In the limit of large reactant concentration, we may assume that the solute release   occurs  at a fixed rate, or {activity}, $A(x)$,
\begin{equation}\label{eq:chemBC}
\left.D\pard{C}{y}\right|_{y=0}=-A(x),
\end{equation} 
where    $D$  is the molecular diffusivity of the solute. 
At sufficiently small   length scales, both advective and unsteady transports are negligible and the  dynamics of the solute concentration is purely diffusive, $D\nabla^2C=0$. For simplicity, we assume that   the upper wall  allows for free exchanges of solute with a chemical reservoir so that the product concentration along it is homogeneous, $C(y=H)=C_0$. These conditions uniquely determine the solute concentration  within the channel as
\begin{equation}
C(x,y) =C_0- \frac{L}{2\pi D}\sum_{n=-\infty}^\infty \frac{a_n}{n}\frac{\sinh\left[ \frac{2n\pi (y-H)}{L}\right]\ee^{\frac{2\ci\pi nx}{L}}}{\cosh\left[ \frac{2n\pi H}{L}\right]} ,\label{eq:csol}
\end{equation}
with $a_n$ the Fourier coefficients of $A(x)$ \change{given by}
\begin{equation}
a_n=\frac{1}{L}\int_0^LA(x)\exp\left(-\frac{2\ci n\pi x}{L}\right).\label{eq:fourierA}
\end{equation}

 Due to the differential affinity of the chemically-patterned wall with the reactant and product molecules, local surface gradients in solute  concentration result in a net slip velocity outside a thin interaction layer providing an effective slip boundary condition for the flow velocity,  $\ub$, as~\cite{anderson1989}
\begin{equation}
\left.( \ub\cdot{\bf e}_x)\right|_{y=0}=\left.M(x)\pard{C}{x}\right|_{y=0}\label{eq:slipbc},
\end{equation}
with $M(x)$ the local diffusiophoretic mobility along the wall. \change{This mobility stems from the difference in affinity with the wall surface  (or short-range interaction potential) between the solute and solvent molecules within a thin interaction layer~\cite{anderson1989}.} 
This simple framework can be easily generalized to other phoretic mechanism, such as thermophoresis~\cite{anderson1989,bickel2013,baraban2013}, or other geometries (e.g.~axisymmetric channels or patterning of top and bottom walls).

Using the fundamental properties of Stokes' flow, the resulting flow rate induced by the phoretic pump through any cross-section $\mathcal{S}_x$ of the channel is 
\change{given by}~\cite{michelin2015c} 
\begin{equation}
Q=\int_{S_x}\ub\cdot\mathbf{\dd S}=\frac{H}{2}\langle u_s(x)\rangle,\label{eq:ratedef}
\end{equation}
where $\langle \cdot\rangle$ is the average in $x$ over a period $[0,L]$. Using Eqs~\eqref{eq:csol}, \eqref{eq:slipbc} and \eqref{eq:ratedef}, the pumping rate $Q$ is then obtained   as
\begin{equation}
Q=-\frac{ H}{D}\sum_{n=1}^\infty \tanh \left(\frac{2n\pi H}{L}\right)  \mbox{Im}[a_nm_{-n}],\label{eq:qgen}
\end{equation}
where $m_n$ are the Fourier coefficients of the mobility $M(x)$.

As expected for phoretic problems in the diffusive limit, the pumping rate is a bilinear function of the activity, $A(x)$, and mobility, $M(x)$, and no pumping is possible if either is constant along the channel, nor if  $M(x)=\lambda A(x)+\mu$ where $\lambda$ and $\mu$ are two arbitrary constants. 
For a Janus-type channel patterning consisting of the repetition of  two patches with properties   $(A_1,M_1)$ and $(A_2,M_2)$, one can write 
\begin{equation}
A(x)=\frac{A_2M_1-A_1M_2}{M_1-M_2}+\left(\frac{A_1-A_2}{M_1-M_2}\right)M(x).
\end{equation}
Consequently,  two-patch patterns are unable to pump, a result which was  expected since such systems are left-right symmetric (i.e.~$x\leftrightarrow -x$) with respect to the midpoint of any of the patches.  This is of course a fundamental difference with phoretic propulsion of microparticles for which a  minimal two-patched Janus patterning leads in general to locomotion~\cite{golestanian2007}. 

\subsection{Pumping rate of a $P$-patch channel}
While realizing continuous variations of the chemical properties of the wall is experimentally difficult,   a simple  patterning    consists of the periodic repetition of $P\geq 3$ patches: on each patch $S_p$ of length $L_p$ (with $\sum_{j=1}^P L_j=L$), both  $A(x)$ and $M(x)$ are constant and take   values $A_p$ and $M_p$, i.e.
\begin{equation}
A(x)=\sum_{p=1}^PA_p\mathbf{1}_{S_p}(x),\qquad  M(x)=\sum_{p=1}^PM_p\mathbf{1}_{S_p}(x),
\end{equation} 
where   $\mathbf{1}_{S_p}(x)=1$ for $x\in S_p$ and $\change{\mathbf{1}_{S_p}(x)}=0$ otherwise.  The Fourier coefficients $a_n$ and $m_n$ can  be obtained  from Eq.~\eqref{eq:fourierA}, and the flow rate of the channel is computed from Eq.~\eqref{eq:qgen} as 
\begin{align}
Q/L=\sum_{n=1}^\infty&\frac{h\,\tanh(2\pi nh)}{\pi^2n^2}\sum_{p<q}\alpha_{pq}\sin\left(\pi n l_p\right)\sin\left(\pi n l_q\right)\sin\left(\pi n \left[l_p+2\displaystyle\sum_{j=p+1}^{q-1}l_j+l_q\right]\right),\label{eq:qstripe}
\end{align}
with $l_p=L_p/L$ the reduced length of $S_p$, $h=H/L$ the channel aspect ratio and $\alpha_{pq}=(M_pA_q-M_qA_p)/D$. This generic form expresses the pumping rate in the channel  as the sum of   pair interactions between patches, whose intensity depends on their  lengths and the distance between their centers. Note that the flow rate $Q$ depends on the $P(P-1)/2$ coefficients $\alpha_{pq}=-\alpha_{qp}$ rather than the $2P$ chemical characteristics $(A_j,M_j)$ but $\alpha_{pq}$ may not   be defined independently from each other. \change{These coefficients also set the characteristic velocity scales generated in such pumps, which are similar to those for the flows generated by phoretic swimmers~\cite{golestanian2007}.}

\subsection{The optimal and minimal phoretic pump}
Since channels with $P=2$ can never pump, the minimal phoretic pump has $P=3$ patches.  In that case,  using $\sum l_j=1$, the pumping rate in Eq.~\eqref{eq:qstripe} becomes~\cite{SM}
\begin{align}
\label{finalQ}{Q}/{L}&=(\alpha_{12}+\alpha_{23}+\alpha_{31})\times\mathcal{G}(l_1,l_2,l_3,h),\qquad \textrm{with}\quad
\mathcal{G}(l_1,l_2,l_3,h)=\sum_{n=1}^\infty\frac{h(-1)^{n+1}\tanh(2\pi nh)}{\pi^2n^2}\prod_{j=1}^3\sin\left(\pi n l_j\right),%\label{eq:q3}
\end{align}
and is   the product of two functions: (i) $\mathcal{F}(A,M)=\alpha_{12}+\alpha_{23}+\alpha_{31}$ which depends exclusively on the chemical properties of the   patches and (ii) $\mathcal{G}(l_1,l_2,l_3,h)$ which depends only on the  geometry of both channel and  patches. Note that the function $\mathcal{G}$, written here in a symmetric form with respect to $(l_i)_{1\leq i\leq 3}$, is effectively a function of $l_1$, $l_2$ and $h$ only (since $l_1+l_2+l_3=1$).
   
The chemical function $\mathcal{F}$ can be rewritten (using the convention $A_{j+3}=A_j$), 
\begin{equation}
\mathcal{F}(A,M)=\sum_{j=1}^3(A_{j+1}-A_{j})(M_j+M_{j+1}),
\end{equation} 
and can  thus be interpreted as the sum of contributions of adjacent pairs of patches which  each induces a net flow proportional to the mean mobility multiplied by the difference in activity. A similar  result is  at the heart of the self-propulsion of Janus microswimmers~\cite{golestanian2007}.

The explicit separation of the chemical and geometric dependences of the pumping rate in Eq.~\eqref{finalQ} confers a universality to the three-patch configuration: The variation of the  flow rate with the geometric patterning of the channel is not affected by the values of the chemical activities or mobilities. In particular, this means that the optimal pump, found by   maximising   the function $\mathcal{G}$, is unique and identical for all chemistry. 
\begin{figure}[t]
\begin{center}
%\begin{tabular}{cc}
%\multicolumn{2}{c}{\subfigure[~Representation definition]{\includegraphics[width=.5\textwidth]{Figure2.pdf}} }\\
%\subfigure[~$h=0.5$]
%{\includegraphics[width=.3\textwidth]{G_h05.pdf}} &
%\quad\quad\subfigure[~$h=2$]
%{\includegraphics[width=.3\textwidth]{G_h2.pdf}} 
%\end{tabular}
\includegraphics[width=.65\textwidth]{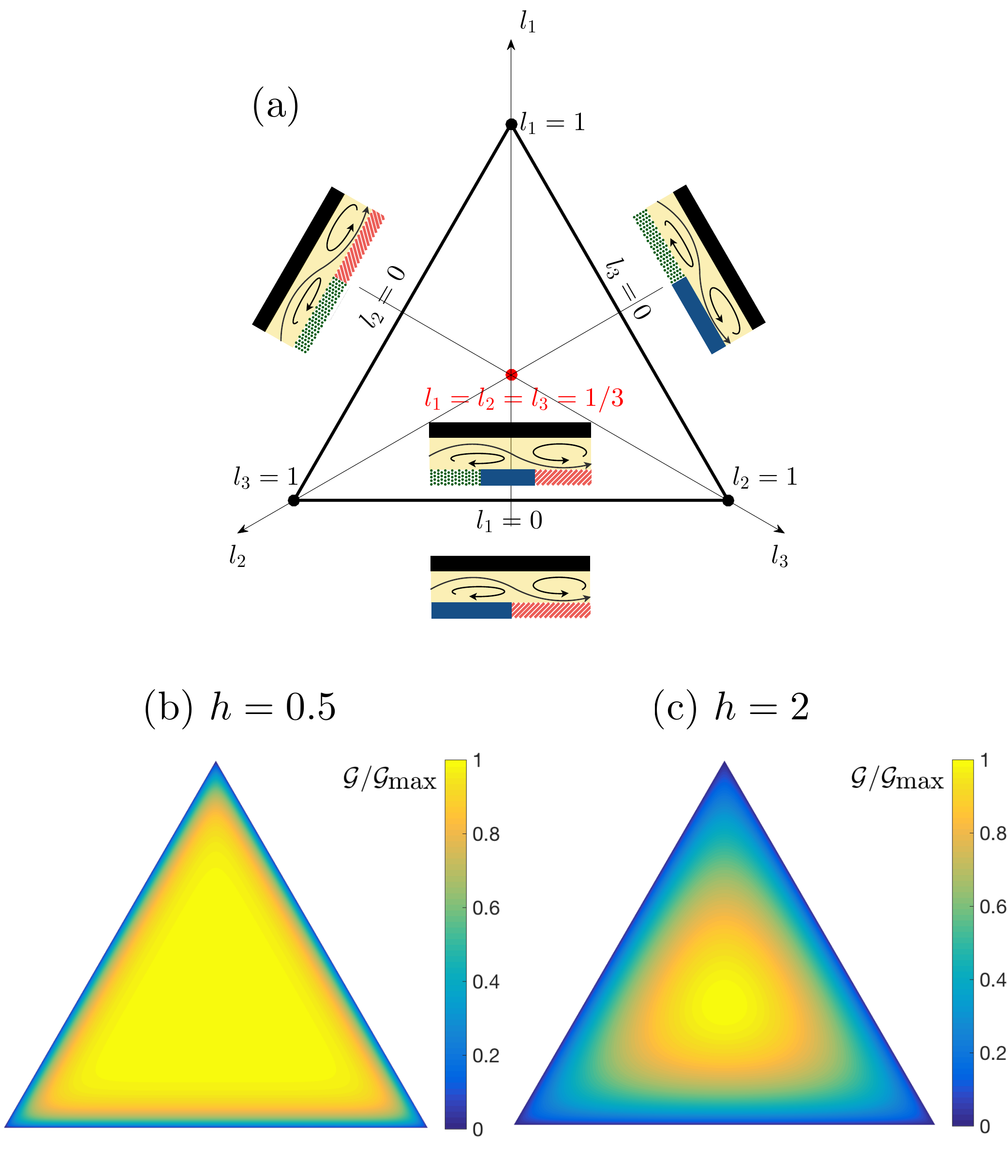}
\caption{\textbf{Influence of wall patterning on the performance of minimal 3-patch phoretic pumps}. (a): Parametric representation of the geometric configuration $(l_1,l_2,l_3)$ with $l_1+l_2+l_3=1$ (\change{flow illustration is schematic}). (b,c): Iso-values of the function  $\mathcal{G}$ quantifying the contribution of the  geometry of chemical patches to phoretic pumping for $h=0.5$  (b) and  $h=2$ (c).
}\label{fig:geometry}
\end{center}
\end{figure}  

The variation  of $\mathcal{G}$ within the $2D$ parameter space $\mathcal{I}_3=\left\{0\leq l_1,l_2,l_3\leq 1,\,\sum l_i=1\right\}$  is shown in Fig.~\ref{fig:geometry}. For all aspect ratios $h$, $\mathcal{G}$ vanishes if any $l_j=0$ (Janus limit), which are the boundary points on 
 $\mathcal{I}_3$. The  gradient of $\mathcal{G}$  with respect to $(l_1,l_2)$ is given by
\begin{align}
\left(\pard{\cal G}{l_1}\right)_{l_2}&=\sum_{n=1}^\infty\frac{h\tanh(2\pi nh)}{n\pi}\sin(n\pi l_2)\sin[n\pi(l_1-l_3)],
%\\ \left(\pard{\cal G}{l_2}\right)_{l_1}&=\sum_{n=1}^\infty\frac{h\tanh(2\pi nh)}{n\pi}\sin(n\pi l_1)\sin[n\pi(l_2-l_3)],
\end{align}
with $l_3=1-l_1-l_2$ and $(\partial{\cal G}/\partial l_2)$ is obtained similarly; the only point within $\mathcal{I}_3$ where $|\cal G|$ has a maximum is $l_1=l_2=l_3=1/3$, which confirms the results of Fig.~\ref{fig:geometry}. 
The optimal minimal (3-patch) pump is therefore unique and, independently of the chemistry, is the one where all patches have equal lengths.

\begin{figure}
\begin{center}
\includegraphics[width=.5\textwidth]{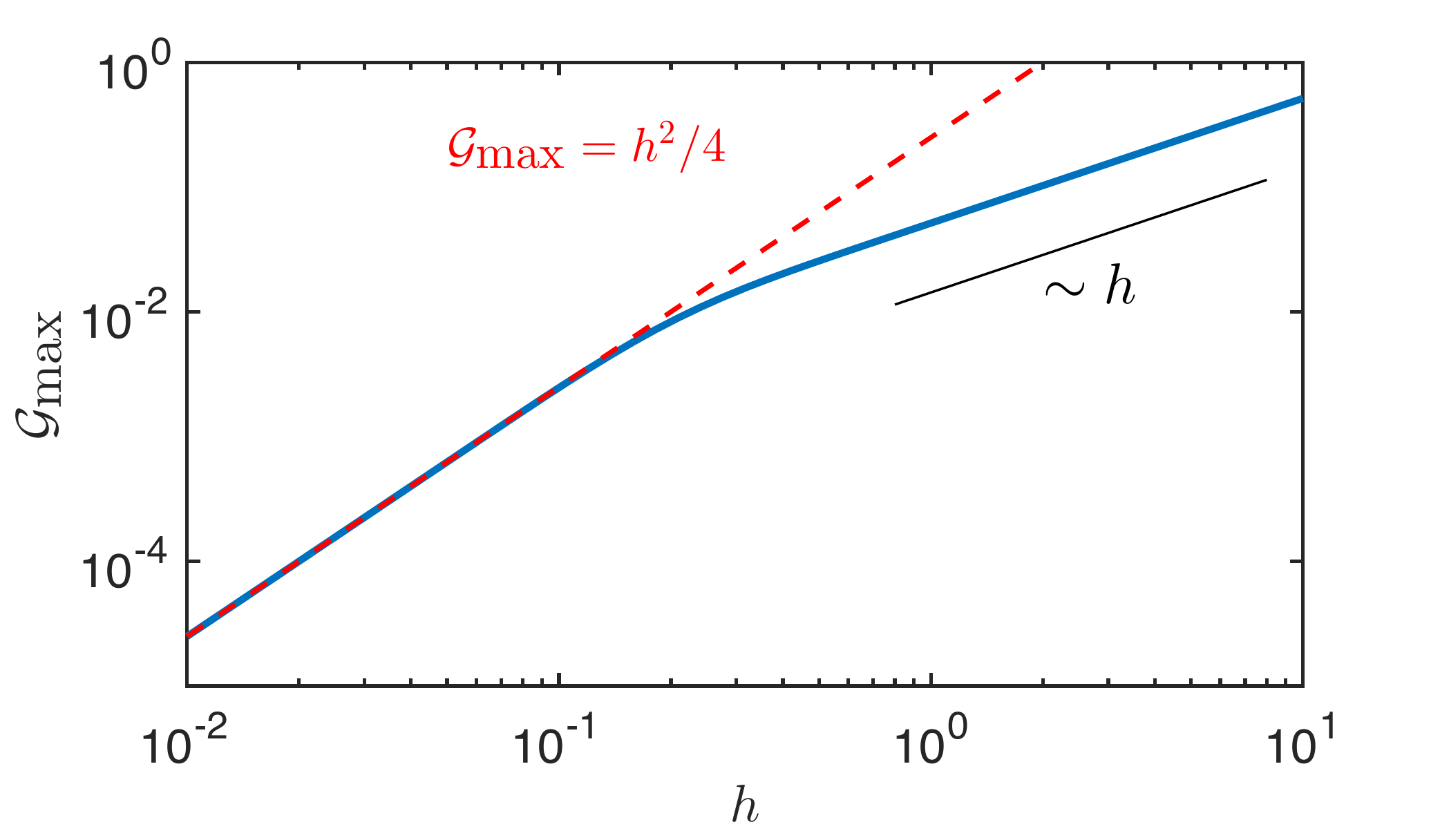}
\caption{\textbf{Influence of the pump aspect ratio on the performance of minimal $3$-patch phoretic pumps}. The dependence of the  maximum pumping rate, ${\cal G}_{\rm max}$, on the channel aspect ratio $h=H/L$ is shown as well as the asymptotic prediction for $h\ll 1$ (red).}\label{fig:Gmax}
\end{center}
\end{figure}

The dependence of the pumping ability of the channel on its geometry can be further understood by examining the impact of the channel aspect ratio, $h=H/L$, on the optimal flow rate, ${\cal G}_{\rm max}$ (Fig.~\ref{fig:Gmax}). For large $h$, the flow rate varies linearly with $h$. In that case,     the concentration distribution at the lower walls is independent of $h$ at leading order, except for its mean value which does not contribute to pumping. The resulting phoretic slip forcing is therefore independent of $h$, and similarly to Couette (shear) flow, the net pumping is linear in $h$.

In the opposite limit, $h\ll 1$, the maximum flow rate scales quadratically  with $h$ (Fig.~\ref{fig:Gmax}). In that case, the concentration profile is almost piecewise constant in $x$ (i.e.~away from the junctions between patches).  
Zooming in on the boundary between patches $j$ and $j+1$ for $x\approx x_j$, the leading-order concentration $c$ can be rewritten as
\begin{equation}
C=C_0+\frac{(A_j+A_{j+1})(H-y)}{2}+\frac{(A_j-A_{j+1})}{2}\tilde{C}\left(\frac{x-x_j}{H}\right),
\end{equation}
where $\tilde{C}(s)$ is an odd function of $s$ with $\tilde{C}(\pm\infty)\pm 1$. Since $M(x)$ is piecewise constant   near $x_j$, the resulting contribution of this junction to the pumping flow rate, $Q_{j,j+1}$, is obtained at leading-order for $h\ll 1$ as
\begin{align}
%Q_{j,j+1}&=\frac{HM_j}{2DL}\int_{-l}^0\left.\pard{c}{x}\right|_{y=0}\dd x+\frac{HM_{j+1}}{2DL}\int_{0}^l\left.\pard{c}{x}\right|_{y=0}\dd x\nonumber\\
Q_{j,j+1}&=\frac{L(A_{j+1}-A_{j})(M_j+M_{j+1})h^2}{4D}\cdot
\end{align}
The total flow rate thus depends  only on the junction between adjacent patches and  is then obtained (for  $P$ patches) as 
\begin{equation}
Q/L\sim\frac{h^2}{4D}\sum_{j=1}^P(A_{j+1}-A_{j})(M_j+M_{j+1})=\frac{h^2}{4}\sum_{j=1}^P\alpha_{j,j+1}.
\end{equation}
 Comparing   with Eq.~\eqref{finalQ} for $P=3$ shows that $\mathcal{G}_\textrm{max}= h^2/4$,   in excellent agreement with the full solution (Fig.~\ref{fig:Gmax}), and that $\mathcal{G}\approx\mathcal{G}_\textrm{max}$ when $h\ll 1$, as also observed on Fig.~\ref{fig:geometry}, demonstrating the robustness of the optimal design in that limit.

The universal nature of the optimal geometry for minimal ($P=3$) phoretic pumps 
  is intimately linked to the number of independent   chemical properties setting the flow rate. For $P$ patches, $2P$ different  properties come into play, ($A_i,M_i$). Denoting by  $\mathcal{A}$ and $\mathcal{M}$ a characteristic magnitude of  activity and mobility, dimensional analysis imposes that $Q=\mathcal{AM}\times\tilde{Q}$, and $\tilde Q$ depends on only $2(P-1)$ parameters. No net pumping is obtained if either $(A_i)_i$ or $(M_i)_i$ are all identical, or if both sets are linearly correlated, providing three additional constraints, such that the pumping rate effectively only depends on $2P-5$ independent chemical parameters. For $P=3$ patches, this confirms that a single chemical function controls the pumping rate, conferring its universality to the minimal pump.

\subsection{Minimal phoretic swimmers vs. minimal phoretic pumps}

While the minimal phoretic pump must include three different patches,   minimal phoretic swimmers are able to break symmetries using   only two. However, in contrast to the results obtained above and showing universality of the three-patch pump,  the optimal minimal (Janus) swimmer is not universal but its geometry depends on the surface chemistry. This can be seen by evaluating  the swimming velocity of an unit-radius axisymmetric Janus sphere coated with two different materials $(A_1,M_1)$ on the portion $\mu\leq z\leq 1$ of its surface (front side) and $(A_2,M_2)$ for $-1\leq z\leq \mu$ (back). The result is
\begin{align}
U&=\frac{(A_2-A_1)(1-\mu^2)}{8}\Big[M_1+M_2+(M_2-M_1)V(\mu)\Big],\qquad \textrm{with}\quad V(\mu)=\mu^3+2\sum_{n=2}^\infty\frac{(1-\mu^2)L_n'}{n(n+1)}\left(\frac{L_{n-1}'}{n}-\frac{L_{n+1}'}{n+2}\right),
\end{align}
where $L_n'(\mu)$ is the derivative of the $n$-th Legendre polynomial~\cite{golestanian2007}. The non-universality of Janus swimmers can then be demonstrated by highlighting a few examples. When $M_1=M_2$, the  Janus swimmer with maximum speed is hemispheric, and thus the optimal value is $\mu_\textrm{opt}=0$. In contrast, when $M_1=-M_2$, the hemispheric particle with $\mu=0$ does not swim and instead maximizes its swimming speed for  $\mu_\textrm{opt}\approx\pm 0.61$. The optimal Janus swimmer, i.e.~the value of $\mu$ maximizing $|U|$, is therefore not universal and optimizing the patterning of the surface of the swimmer  requires a detailed knowledge of its chemical properties, in contrast with  minimal phoretic pumps which are always optimal for $l_i=1/3$.

\subsection{Optimal pumps beyond $3$-patch patterns}
\begin{figure}[h!]
\begin{center}
\includegraphics[width=.65\textwidth]{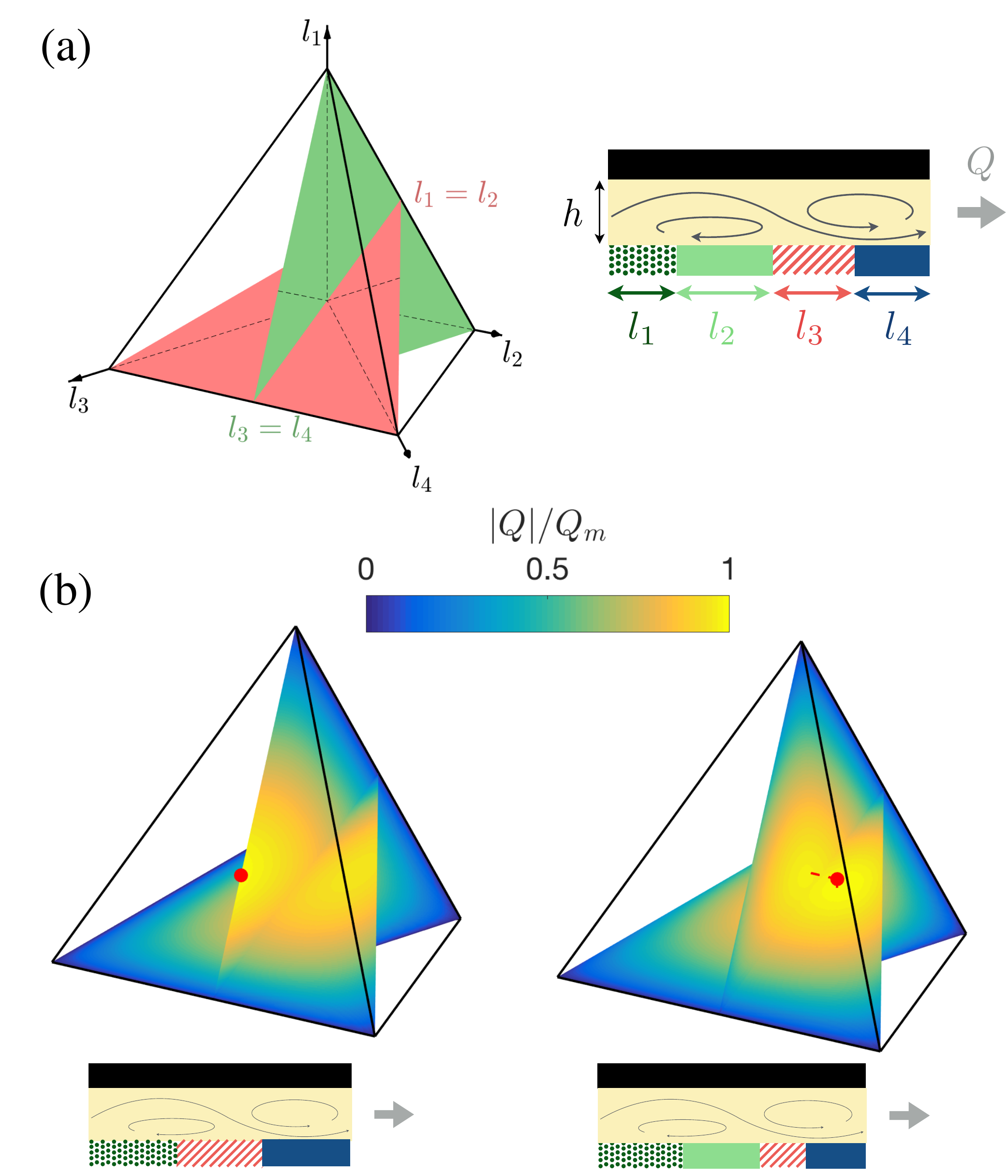}
\caption{\textbf{Optimal $4$-patch phoretic pumps} (a) Parametric representation of the four-patch pump where each $l_i$ is measured along a height of a regular tetrahedron. (b) The evolution with $(l_i)$ of the flow rate  visualized for $l_1=l_2$ and $l_3=l_4$ and two different fixed sets of chemical properties, one leading to an optimal degenerated pump (only three patches, left) and one with four different patch  lengths (right). In each case, the parametric position of the optimal configuration is also shown (red point) together with the structure of the optimal pump. \change{Degenerate pumps with only three patches correspond in this representation to the four faces of the tetrahedron, and as such the planar representation of Fig.~\ref{fig:geometry} is simply the projection of the present figures on the particular subspace of interest.}}\label{fig:4stripe}
\end{center}
\end{figure}
The universality for pumps is lost for $P>3$ as the pumping rate now depends on $2P-5>1$ independent chemical parameters. In the case of $P=4$ patches, the net flow rate, Eq.~\eqref{eq:qstripe}, becomes
\begin{align}
&Q/L=\sum_{j=1}^4\mathcal{F}_j(A,M)\mathcal{H}_j(l_1,l_2,l_3,l_4,h),\quad \textrm{with  }\label{eq:4nus}\\
&\mathcal{F}_1=\alpha_{23}+\alpha_{34}+\alpha_{42},\qquad\mathcal{H}_1=\sum_{n=1}^\infty\frac{h(-1)^{n+1}\tanh(2\pi nh)}{\pi^2n^2}\cos(\pi nl_1)\prod_{k=2}^4\sin(\pi nl_k).
\end{align}
with $\mathcal{F}_j$ and $\mathcal{H}_j$   obtained by circular permutation for $j\geq 2$~\cite{SM}. The $\mathcal{F}_j$ contribution is essentially a modulation of the $3$-patch pump obtained for $l_j=0$. The pumping rate nevertheless depends on only three independent parameters since these four contributions are not independent ($\mathcal{F}_1+\mathcal{F}_3=\mathcal{F}_2+\mathcal{F}_4$). 
All possible geometries now span the three-dimensional parameter space $\mathcal{I}_4=\left\{0\leq l_1,l_2,l_3,l_4\leq 1,\,\sum l_i=1\right\}$. Depending on the surface chemistry, the optimal pumping rate is reached (i) within $\mathcal{I}_4$ if $\mathcal{F}_1\mathcal{F}_3$ and $\mathcal{F}_2\mathcal{F}_4$ are both positive (non-trivial $4$-patch pump) or (ii) on its boundary if either quantity is negative (degenerated $3$-patch pump), in which case the universal optimal pump with three equal-length patches is recovered~\cite{SM}. \change{These two possibilities are illustrated on Fig.~\ref{fig:4stripe} where the dependence of the pumping efficiency on the  geometry of patterning is represented over $\mathcal{I}_4$.}

\section{Discussion}
In summary, this work proposes a generic mathematical framework to evaluate and optimize the phoretic pumping performance of a straight microchannel periodically-coated with active surfaces. Focusing on patterns which are well  suited for experimental realization, namely a succession of materials with uniform chemical properties (patches), we show that the minimal pump  features three different patches and   is optimal for three patches of equal lengths regardless of their   chemical properties.  Although we focused on  diffusiophoresis, our results  are also  applicable to thermophoresis and electrophoresis (at least in the  weak gradient limit \change{when surface slip is proportional to the concentration gradient~\cite{anderson1989}}) and could be  extended to more complex geometries using numerical computations. \change{For clarity and generality, we purposely focused here on the simplest chemical formulation of the problem, i.e.~a prescribed fixed-flux of a single chemical component (reactant or product). Our  framework could nevertheless  be extended to account for a more detailed description of the chemical reaction, for example by including several chemical components or multi-step reactions to describe the wall activity.}

The most important result of our study is the universality of the optimal geometric design. This is  a unique feature of the phoretic pumping problem that does not have an equivalent for phoretic swimmers. Furthermore, this universality is likely to be critical for  experimental development since determining independently the chemical and phoretic properties of active materials is challenging experimentally. There is therefore  no need for a trial-and-error experimental approach to phoretic pumps.

\begin{acknowledgements}
This project has received funding from the European Research Council (ERC) under the European Union's Horizon 2020 research and innovation programme  (grant agreements 714027 to SM and 682754 to EL).
\end{acknowledgements}

%\bibliography{pumps.bib}
%\bibliographystyle{unsrt}
\end{document}